\documentclass[aps,twocolumn,amsmath,amssymb,showpacs,prb]{revtex4}
\usepackage{epsfig}

\begin{document}

\begin{abstract}
We present an NMR study of Na$_2$C$_{60}$ and K$_4$C$_{60}$, two
compounds that are related by electron-hole symmetry in the
C$_{60}$ triply degenerate conduction band. In both systems, it is
known that NMR spin-lattice relaxation rate (1/T$_1$) measurements
detect a gap in the electronic structure, most likely related to
singlet-triplet excitations of the Jahn-Teller distorted (JTD)
C$_{60}^{2-}$ or C$_{60}^{4-}$. However, the extended temperature
range of the measurements presented here (10 K to 700 K) allows to
reveal deviations with respect to this general trend, both at high
and low temperatures. Above room temperature, 1/T$_1$ deviates
from the activated law that one would expect from the presence of
the gap and saturates. In the same temperature range, a lowering
of symmetry is detected in Na$_2$C$_{60}$ by the appearance of
quadrupole effects on the $^{23}$Na spectra. In K$_4$C$_{60}$,
modifications of the $^{13}$C spectra lineshapes also indicate a
structural modification. We discuss this high temperature
deviation in terms of a coupling between JTD and local symmetry.
At low temperatures, 1/T$_1$T tends to a constant value for
Na$_2$C$_{60}$, both for $^{13}$C and $^{23}$Na NMR. This
indicates a residual metallic character, which emphasizes the
proximity of metallic and insulting behaviors in alkali
fullerides.
\end{abstract}

\author{V. Brouet, H. Alloul }

\affiliation{Laboratoire de Physique des Solides, Universite
Paris-Sud, Bat 510 91405 Orsay (France) }

\author{S. Garaj, L. Forr\'o}
\affiliation{ Laboratoire des solides semicristallins,
IGA-Departement de Physique, Ecole Polytechnique Federale de
Lausanne, 1015 Lausanne (switzerland)}

\title{Gaps and excitations in fullerides with partially filled
bands :\\
NMR study of Na$_2$C$_{60}$ and K$_4$C$_{60}$}

\date{\today} \newpage
\maketitle
\section{Introduction}

Early after the discovery of fullerides, A$_4$C$_{60}$ has been found to
behave as an insulator rather than the metal expected in a band picture \cite
{Erwin}. The first indication for this was the detection by $\mu $SR in K$_4$%
C$_{60}$ of a muonium precession \cite{KieflPRL92} at 5K, which is
known to be quickly suppressed in presence of unpaired electrons.
This signal disappears at higher temperatures suggesting the
presence of thermally populated states. This has been confirmed by
subsequent measurements of magnetic properties, ESR detects an
activated susceptibility in K$_4$C$_{60}$ with E$_a=$ 60 meV
\cite{Petit} and SQUID in Rb$_4$C$_{60}$ yields E$_a=60$ meV
\cite{SQUID}. An activated component has also been found in NMR
1/T$_1$
in K$_4$C$_{60}$ with E$_a=55$ meV \cite{ZimmerK4} and Rb$_4$C$_{60}$ with E$%
_a=70$ meV \cite{ZimmerPRB95,Kerkoud}. Other measurements also suggest an
insulating ground state, no Fermi edge is visible by photoemission in K$_4$C%
$_{60}$ and Rb$_4$C$_{60}$ films \cite{HesperPRB2000} and no Drude
peak is found by optical conductivity \cite{IwasaPRB95}. However,
in this latter study, the gap to the lowest conductivity peak is
significantly larger than in magnetic measurements, around 500
meV. A more recent investigation by EELS\ in transmission
\cite{Fink} also revealed a gap of the order of 500 meV in
K$_4$C$_{60}$ and Rb$_4$C$_{60}.$ Therefore, two different gaps
are necessary to describe these systems, a small ``spin-gap'' of
the order of 50-100 meV and a larger ``optical'' gap of about 500
meV. As no magnetism has ever been reported at low temperature,
the ground state for C$_{60}^{4-}$ must be singlet and the small
gap has been associated to singlet-triplet transitions
\cite{ZimmerPRB95,Kerkoud,BrouetPRL2001}. The large gap is
presumably a direct gap in the substructure of the band. In
addition, these systems are close to a metal-insulator transition,
as shown by the transition to a metallic state observed by NMR at
12 kbar in Rb$_4$C$_{60}$
\cite{Kerkoud} and by the fact that the system with the smallest C$_{60}$-C$%
_{60}$ distance, Na$_4$C$_{60},$ might even be metallic in its monomer phase
(T 
\mbox{$>$}%
500K), namely a body centered tetragonal ($bct$) structure \cite
{ZhouCox,Bendele}, isostructural to other A$_4$C$_{60}$ systems
\cite {OszlanyiPRB98}.

Understanding the origin of the insulating state is a necessary step to
describe the physics of fullerides. Although the crystal field due to the
bct structure is not sufficient to lift the threefold degeneracy of the t$%
_{1u}$ band (the C$_{60}$ lowest unoccupied molecular orbital,
filled with electrons brought by the alkali ions) \cite{Erwin}, it
has been suggested that this particular structure might explain
why A$_4$C$_{60}$ is insulating contrary to the metallic and face
cubic centered ($fcc$) A$_3$C$_{60}$ compounds
\cite{GunnarssonA4}. Especially, the $bct$ lattice is bipartite
(contrary to the $fcc$ one) which could play a role by enhancing
antiferromagnetic correlations, hence favoring a Mott insulating
state. However, we have shown recently that Na$_2$C$_{60}$, which
is cubic \cite{Yldirim}, exhibits a behavior similar to
A$_4$C$_{60},$ with an activated temperature dependence of the NMR
1/T$_1$ \cite{BrouetPRL2001}. Electron-hole symmetry then applies
in the t$_{1u}$ band and the detection of a gap is an intrinsic
feature of fullerides with 2 or 4 electrons per ball rather than
one of the bct structure. The discrepancy with band structures
must be sought in an underestimation of the role of
electron-electron and/or electron-phonon interactions. The fact
that these two interactions are strong and have similar orders of
magnitude is actually one of the most interesting aspect of the
physics of fullerides. This paper is a first of a serie of three
papers (called hereafter I , II \cite{BrouetPartII} and III \cite
{BrouetPartIII}), whose purpose is to present an extensive NMR
study of different stoichiometries to clarify this point.

\begin{figure}[tb]
\centerline{ \epsfxsize=0.45\textwidth{\epsfbox{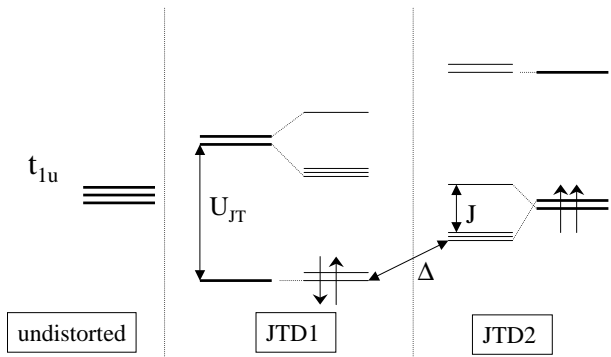}} }
\caption{ Schematic representation of the structure of the three
t$_{1u}$ levels for two electrons per C$_{60}$ and for the two
most stable distortions (called JTD~1 and JTD~2) without spin
degeneracy (thick lines) and with spin degeneracy (thin lines).
The various gaps are indicated by arrows, the Jahn-Teller
splitting U$_{JT}$, the exchange splitting J and the
singlet-triplet gap $\Delta $ (adapted from ref.~\cite{Manini}). }
\label{molecule}
\end{figure}
There is a growing consensus that Jahn-Teller distortions (JTD) of the C$%
_{60}$ molecule are an essential ingredient for the insulating state of A$_4$%
C$_{60}$, although they have never been detected directly in these
compounds \cite{Bendele}. Molecular calculations \cite{Manini}
indicate that the two most stable JTD correspond to the squeezing
or elongation of the C$_{60}$ molecule around one of the 3
equivalent axes of the quasi-spherical structure. This lifts
the degeneracy of the t$_{1u}$ levels by an amount of U$%
_{JT}\approx 0.5$ $eV$, as sketched on Fig. \ref{molecule}. The
most stable JTD corresponds to a singlet ground state for
C$_{60}^{2-}$ (see the case JTD 1 on Fig. \ref{molecule}) and
C$_{60}^{4-}$ (case JTD 2), as observed experimentally. The first
excited state of a C$_{60}^{2-}$ is a triplet corresponding to JTD
2. It lies about $\Delta =$100~meV higher in energy, which is
consistent with the experimentally measured ``spin-gap''.

In the solid, two different situations could occur, either a
cooperative JT distortion or independent (and possibly dynamic)
JTD for each molecule. In the first case, a band gap could open if
the cooperative JTD is commensurate with the lattice. The second
scenario is considered to be the most likely for fullerides due to
the unusually large quantum fluctuations on the C$_{60} $ molecule
\cite{Fabrizio,Capone}. Another kind of excitation could take
place between the t$_{1u}$ levels split by the JTD, and
U$_{JT}=$~500 meV would correspond to the large ``direct'' optical
gap. Let us emphasize that as $U_{JT}$ has the same order of
magnitude as the band width $W$, so that the gap in other
directions should be much smaller. This is why Fabrizio {\it et
al.} have proposed that strong electronic correlation combined
with JTD are necessary to understand the insulating state, which
could be called a ``Mott Jahn-Teller'' ground state
\cite{Fabrizio}. As we have seen, this model
is supported by the large body of experiments in Na$_2$C$_{60}$ and A$_4$C$%
_{60}$, because it explains the occurrence of two gaps and predicts correct
order of magnitude for them.

In this paper, we will first introduce the basic NMR\ facts
indicating that Na$_2$C$_{60}$ and K$_4$C$_{60}$ are insulators
with similar properties and in good agreement with the ``Mott JT
model'' described previously (section II). We then present data up
to very high temperatures (700 K) that call for a more refined
model than a simple thermal population of an excited state. We
suggest in section III that subtle changes in local symmetry
detected by NMR near room temperature could be a relevant
parameter to explain the high temperature evolution of the
physical properties. The structural modifications likely couple to
the JTD and might modify the equilibrium between singlet and
triplet states. In the last section, we will study in details the
question of a possible coexistence of metallic behavior in
Na$_2$C$_{60}$ with the aforementionned molecular excitations. As
explained previously, this is not inconsistent with the ``Mott JT\
scenario'' and the status of JTD in a metallic or semimetallic
environment is actually an important issue to clarify.

\section{A non-magnetic ground state with gapped excitations}

In this section, we show that the $^{13}$C NMR\ spectra shifts and
lineshapes, as well as the dynamic susceptibility monitored by the
spin lattice relaxation rate 1/T$_1,$ exhibit similar features in
Na$_2$C$_{60}$ and K$_4$C$_{60}$. Both compounds are characterized
by a non-magnetic ground state and a gap in their low-energy
excitations. We present new measurements above room temperature
that allow to investigate this behavior in more details.

\subsection{$^{13}$C NMR spectra}

\begin{figure}[tb]
\centerline{ \epsfxsize=0.45\textwidth{\epsfbox{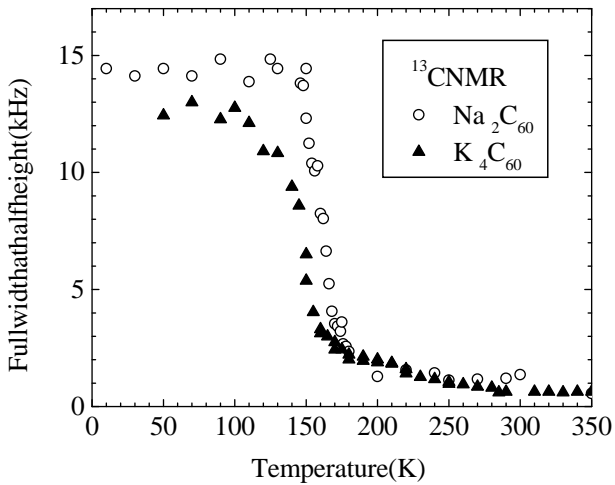}} }
\caption{$^{13}$C NMR linewidth as a function of temperature for
Na$_{2}$C$_{60}$ and K$_{4}$C$_{60}.$ The broadening of the
spectra due to the slowing down of the C$_{60}$ molecular motions
is visible around 150 K. } \label{largeur}
\end{figure}

Let us detail first that the evolution of the $^{13}$C NMR spectra at low
temperatures, shown on Fig. \ref{largeur}, clearly demonstrates that no
magnetic transition occurs in Na$_2$C$_{60}$ and K$_4$C$_{60}$. To
understand this, let us first recall the interactions contributing to the
shift $K$ of one NMR line.

\begin{equation}
\label{k}\overline{\overline{K}}=\overline{\overline{\sigma
}}+\overline{\overline{A}} \text{ }\overline{\overline{\chi }}
\end{equation}

 $\overline{\overline{\sigma }}$ represents a chemical shift
tensor and $ \overline{\overline{A}}$ $\overline{\overline{\chi
}}$ the contribution from unpaired electrons called Knight shift.
$\overline{\overline{A}}$ is the hyperfine coupling tensor between
$^{13}$C and unpaired electrons and $\chi $ the local electronic
susceptibility. All these quantities have both isotropic and
anisotropic parts. In the case of C$_{60}$ compounds, the
anisotropic part is usually the largest because electrons reside
mainly in orbitals with a pronounced $p$ character. As these
orbitals have nodes at
the nuclear position, the electrons do not interact directly with the $^{13}$%
C nuclear spin through the so-called contact interaction and the hyperfine
coupling is mainly of dipolar origin \cite{Pennington}. As a consequence,
the shift is a function of the orientation of one orbital with respect to
the NMR applied field. At high temperatures, when the molecules are rapidly
rotating, the anisotropic contribution is averaged out and narrow lines are
observed. They broaden when the motions slow down and appear static on the
NMR time scale (a few $ms$). This has been observed in many fullerides and
Fig. \ref{largeur} shows that this takes place at 150 K in the case of K$_4$C%
$_{60}$ and 160 K for Na$_2$C$_{60}$. We note that the timescale of the C$%
_{60}$ motion appears to be similar in both compounds, despite the different
structures and presumably different interactions between C$_{60}$ and K or
Na.

There is no further broadening of the spectra below this temperature, which
means that there is {\it no magnetic transition }at least down to the lowest
measured temperature, 10 K in Na$_2$C$_{60}$ and 50 K in K$_4$C$_{60}.$
Indeed, static magnetic moments would create a large local magnetic field on
$^{13}$C nuclei and cause a large broadening of the spectra in these powder
samples. This finding is one key element to involve Jahn-Teller distortions
in the description of these materials because they explain naturally the
singlet ground state. Otherwise, one could have rather expected a magnetic
ground state for localized electrons, because Hund's rule should favor a
high spin state in the t$_{1u}$ levels.

\begin{figure}[tb]
\centerline{ \epsfxsize=0.45\textwidth{\epsfbox{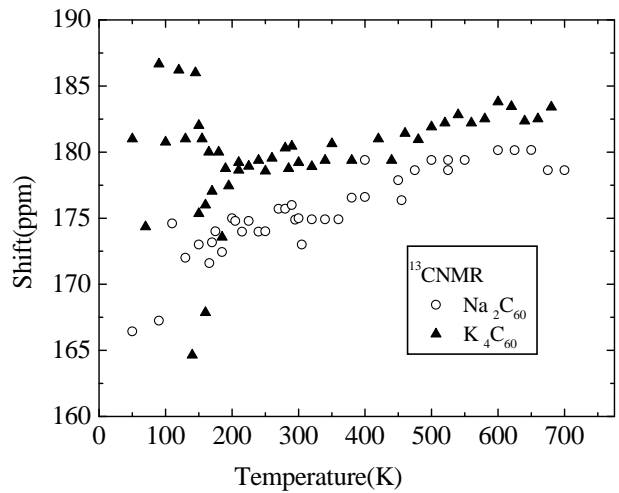}} }
\caption{ Temperature dependence of the shift (with respect to
TMS) of the center of gravity of the $^{13}$C NMR spectra in
Na$_{2}$C$_{60}$ and K$_{4} $C$_{60}.$ The accuracy is lower at
low temperatures because of the broadening of the spectra. }
\label{shift}
\end{figure}

Relevant information about the insulating state should be found in
the temperature dependence of the static spin susceptibility, that
could in principle be extracted from the isotropic part of the
Knight shift. Fig. \ref {shift} shows that this is difficult
because the isotropic shift (defined as the center of gravity of
one spectrum) is small with respect to the linewidth, especially
at low temperature. In pure C$_{60}$, an isotropic chemical shift
$\sigma =143$ ppm is observed \cite{TyckoC60}, characteristic of
the orbital currents flowing in the filled orbitals. This is
expected to be the order of magnitude of the reference for the
Knight shift in alkali fullerides. This idea was confirmed by the
close value of 156 ppm found in the band insulator A$_6$C$_{60}$
\cite{Reichenbach}, which suggests an
empirical correlation for $\sigma $ of +1.5 ppm per added electron. In K$_4$C%
$_{60}$ and Na$_2$C$_{60}$, the shifts are however much larger
than what would be expected with such a contribution for $\sigma $
alone. At room temperature $K=175$ ppm for Na$_2$C$_{60}$ and 179
ppm in K$_4$C$_{60}$, which is comparable to shifts measured in
metallic A$_3$C$_{60}$ (189 ppm in K$_3$C$_{60}$ for example
\cite{TyckoPRL92}). This sizable Knight shift indicates the
presence of electronic excitations giving rise to a large
electronic susceptibility at room temperature.

\begin{figure}[tb]
\centerline{ \epsfxsize=0.45\textwidth{\epsfbox{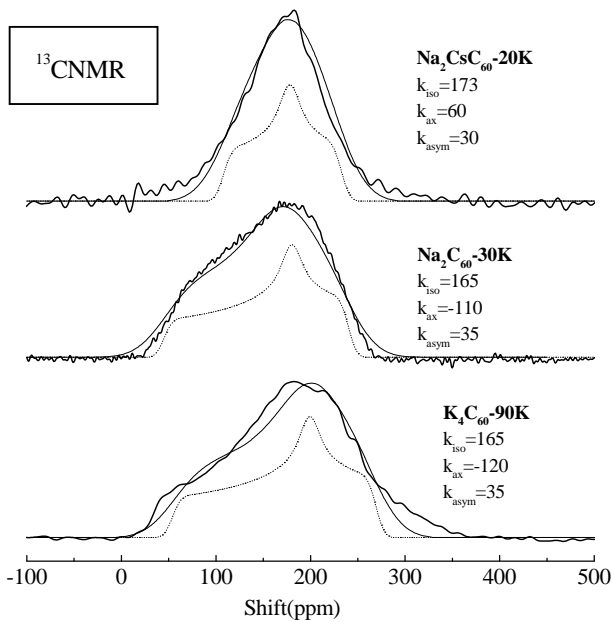}} }
\caption{$^{13}$C NMR spectra at low temperatures for
K$_{4}$C$_{60}$, Na$_{2}$C$_{60} $ and Na$_{2}$CsC$_{60}$ (thick
lines)$.$ Fit to a powder pattern of the anisotropic NMR shift
(Eq.~2) are also presented with parameters indicated on the
figure. The thin solid line is obtained by a convolution of the
theoretical lineshape with a gaussian of half width at half
maximum 35 ppm. A convolution with a 10 ppm large gaussian is also
shown (dotted line) that reveals more clearly the underlying
structure of the spectra. } \label{13CNMR}
\end{figure}

At low temperature, the anisotropic part of the shift, which is also
proportional to the susceptibility, can be studied as well. Typical $^{13}$C
lineshapes are presented on Fig. \ref{13CNMR}, they are similar in Na$_2$C$%
_{60}$ and K$_4$C$_{60}$ with a shoulder on the low frequency side,
characteristic of the chemical shift anisotropy found in pure C$_{60}$ \cite
{TyckoC60}. This can be stated more quantitatively if one extracts the
parameters for the shift tensor by fitting the spectra to the theoretical
powder pattern \cite{Slichter}. Defining,

\begin{equation}\label{tenseur}
K=K_{iso}+K_{ax}\left( \frac{3\cos {}^2\theta -1}2\right)
+K_{asym}\sin {}^2\theta \cos 2\varphi \end{equation}

where $\theta $ and $\varphi $ are the spheric coordinate for the
orientation of the principal axis of the tensor with respect to
the applied magnetic field, one can compute the actual lineshape
by averaging on all possible orientations. The values found for
Na$_2$C$_{60}$ and K$_4$C$_{60}$ are reported on Fig.
\ref{13CNMR}, together with the fit of the experimental spectra. A
convolution with a gaussian function of width 35 ppm has been used
to take into account an experimental broadening and reproduce the
spectra. Because the parameters are not independent, we estimate an error $%
\pm 5$ ppm for each of them. K$_{ax}$ and K$_{asym}$ are very similar to the
parameters found in pure C$_{60}$ (K$_{ax}=-110$ ppm and K$_{asym}=$ 35 $\ $%
ppm \cite{TyckoC60}) eventhough the spectra are shifted by about
20 ppm. In contrast, it has been observed that for metallic
A$_3$C$_{60}$ compounds, the addition of the Knight shift
anisotropy leads to {\it narrower} and more {\it symmetric} lines.
This is illustrated by the spectra in Na$_2$CsC$_{60}$ also shown
on the figure. Therefore, the observation of the typical C$_{60}$
lineshape in Na$_2$C$_{60}$ and K$_4$C$_{60}$ is a sign that the
contribution of conduction electrons is weak at low temperature.
The relatively large value found for the isotropic coupling (K=165
ppm) might be due to a slightly larger value of $\sigma $ than the
one estimated previously.

The susceptibility then increases from a small value at low T to a sizable
one at room temperature, which is consistent with the presence of
singlet-triplet excitations in these compounds proposed in the introduction.
We will see in the course of this paper that we can confirm the presence of
these excitations more accurately using other NMR probes.

\subsection{$^{13}$C NMR spin-lattice relaxation}

\subsubsection{Detection of a gap in the electronic structure}

The most efficient way to detect these excitations is through spin-lattice
relaxation measurements (1/T$_1$), which measures the imaginary part of the
electronic susceptibility, that is, if no particular q dependence is
expected.

\begin{equation}\label{T1}
\frac 1{T_1}=\frac{k_BT}{\hbar}A^2\frac{\chi ^{\prime \prime }(\omega _0\text{%
)}}{\omega _0}
\end{equation}

\begin{figure}[tb]
\centerline{ \epsfxsize=0.45\textwidth{\epsfbox{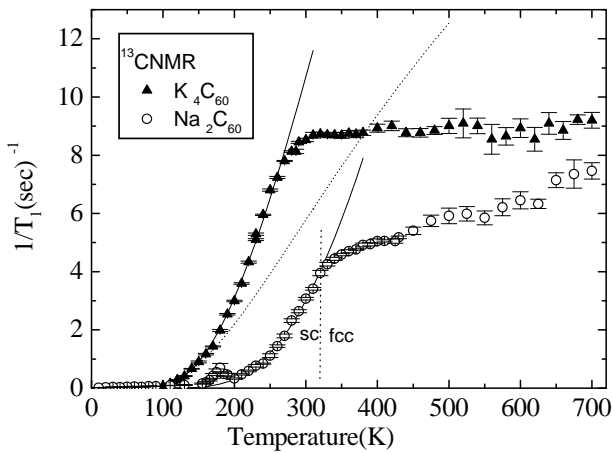}} }
\caption{$^{13}$C NMR 1/T$_{1}$ in Na$_{2}$C$_{60}$ and
K$_{4}$C$_{60}$ from 10 to 700~K.\ Solid lines are fit to an
activated law below room temperature. The dashed line for
K$_{4}$C$_{60}$ is an extrapolation
 of the electronic contribution to 1/T$_{1}$
given in ref.~\cite{ZimmerK4} based on data below 300~K and
assuming a contribution from molecular motion peak. }
\label{T1K4NA2}
\end{figure}

As can be seen on Fig. \ref{T1K4NA2}, 1/T$_1$ increases steeply with
temperature for both compounds, the increase starts around 150~K in K$_4$C$%
_{60}$ and 200 K in Na$_2$C$_{60}$. This has been observed previously (see
ref. \cite{ZimmerK4} for K$_4$C$_{60}$ and ref. \cite{BrouetPRL2001} for Na$%
_2$C$_{60}$) and can be \ attributed to a gap related to
singlet-triplet transitions between two different JTD, as
explained in the introduction.\ The lines on Fig. \ref{T1K4NA2}
correspond to activated laws with E$_g$ = 70~meV for K$_4$C$_{60}$
and E$_g$=140~meV for Na$_2$C$_{60}$ and they describe the data
correctly up to room temperature. At higher temperature,
deviations are observed, which will be discussed in the last
paragraph of this section. In Rb$_4$C$_{60},$ the activated part
of 1/T$_1$ below 250 K is almost quantitatively identical to that
of K$_4$C$_{60}$ \cite{Kerkoud}. This could mean that the 70 meV
gap is characteristic of a JT C$_{60}^{4-}$, while it is nearly
twice larger for C$_{60}^{2-}.$ However, we will argue in this
paper that the gap extracted from 1/T$_1$ could be slightly
different from the molecular value, because it is sensitive to the
details of the local structure. Besides stoichiometry, one
similarity between K$_4$C$_{60}$ and Rb$_4$C$_{60}$ that contrasts
with Na$_2$C$_{60}$ is precisely that they both have a $bct$
structure and this could also be the reason for the different gaps
in Na$_2$C$_{60}$ and A$_4$C$_{60}$.

In a previous study \cite{ZimmerK4}, Zimmer {\it et al.} have
proposed a smaller gap in K$_4$C$_{60}$ (50 meV)$,$ because they
assign part of the increase of 1/T$_1$ to a molecular motion peak.
They suggested that 1/T$_1$ would follow the dashed line of Fig.
\ref{T1K4NA2}, once this peak is substracted. By extending the
measurement to higher temperature, we show
that 1/T$_1$ on the contrary {\it saturates} above 250~K towards 1/T$_1$%
~=~cst, which invalidates this analysis. As for a molecular motion
peak, Fig.~\ref{T1K4NA2} shows that it is unambiguously resolved
at 180~K for Na$_2 $C$_{60}$, but it does not appear clearly for
K$_4$C$_{60}$. In the next section, we study this contribution in
more details, to determine to what extent it could modify the
shape of 1/T$_1$ vs T.

\subsubsection{Realistic parameters for the molecular motion peak
contribution}

Indeed, molecular motions have been found to contribute to 1/T$_1$ in
different fullerides, most notably pure C$_{60}$ \cite{TyckoC60} and K$_3$C$%
_{60}$ \cite{Yoshinari}. This is due to the fact that the local magnetic
field $H_{loc}$ sensed by $^{13}$C nuclear spin depends on the orientation
of the $p_z$ orbital, nearly perpendicular to the C$_{60}$ ball at one
carbon site, with respect to the NMR applied field. Rotation of the ball
will {\it modulate} this local field. If the timescale of the motions is
such that they create fluctuations of H$_{loc}$ at the nuclear Larmor
frequency $\omega _0,$ they can relax the NMR nuclei. For fullerides, the
fast rotation of the C$_{60}$ molecule around one axis can be described by a
frequency 1/$\tau $ that is typically of the order of $\omega _0$ for
temperatures around 200 K- 400 K. A ``Bloembergen Purcell Pound'' peak \cite
{BPP} can be expected in this temperature range with :

\begin{equation}\label{BPP}
 \frac 1{T_1}=\alpha \text{ }(\gamma H_{loc})^2\frac{2\tau
}{1+(\omega _0\tau )^2}
\end{equation}
where $\gamma $ the gyromagnetic ratio for the nuclei and $\alpha $ is a
numeric prefactor of order unity, which depends on the details of the
molecular motion.

\begin{figure}[tb]
\centerline{ \epsfxsize=0.45\textwidth{\epsfbox{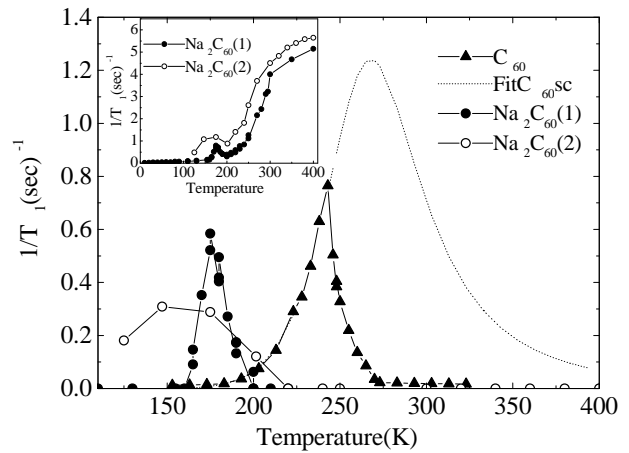}} }
\caption{ Comparison of the molecular motion peak contribution to
1/T$_{1}$ in pure C$_{60}$ (ref.~\protect\cite{TyckoC60}) and two
Na$_{2}$C$_{60}$ samples. The dashed line is a fit of the
molecular motion in the $sc$ phase of C$_{60}$ (T\mbox{$<$}260 K)
given in ref.~\cite{TyckoC60}. Inset : 1/T$_{1}$ as a function of
temperature in the two Na$_{2}$C$_{60}$ samples.} \label{T1motion}
\end{figure}

To determine the actual value of $\alpha ,$ the anisotropy of
H$_{loc},$ measured from the low temperature spectra, should be
used and the molecular motion should be modelled in an appropriate
way, for example a uniaxial rotation along one diagonal axis,
following the lines of \cite{YoshinariPRB}. The maximum value of
$\alpha $ is 1 if the local field is assumed to fluctuate randomly
between two values $\pm H_{loc}$ \cite{Slichter} but it can be
much smaller, for example 6/40 in the case of random molecular
reorientation for an axial symmetry of the shift tensor
\cite{Abragam}. Here, we want to estimate $\alpha $ from the
experiment rather than from a theoretical model; to do so, some
typical molecular contributions to 1/T$_1$ are shown on Fig.
\ref{T1motion}.

From Eq. 4, it can be seen that the maximum of 1/T$_1$ occurs for $\omega
_0\tau =1$ with a value{\it \ depending uniquely on the linewidth $\Delta
\nu =$}$\gamma H_{loc}/2\pi .$ It is given by $1/T_1)_{\max }=\alpha $ $%
(\gamma H_{loc})^2/\omega _0.$ We have shown in the preceding section that K$%
_4$C$_{60}$, Na$_2$C$_{60}$ and C$_{60}$ exhibit roughly the same
linewidth, so that similar contributions should be expected. Pure
C$_{60}$ is the simplest case because the relaxation is dominated
by molecular motions and the field dependence predicted by Eq.~4
has been successfully checked \cite {Johnson}. The data from ref.
\cite{TyckoC60} are reported on Fig. \ref {T1motion}, and yield
$1/T_1)_{\max }=0.8$~sec$^{-1}.$ Let us note that
these authors have extrapolated a ``true'' maximum $1/T_1)_{\max }=1.2$ sec$%
^{-1}$ (dashed line), assuming that the peak is ``cut'' by the orientational
transition at 260 K. This order of magnitude is consistent with the peak $%
1/T_1)_{\max }=0.6\sec {}^{-1}$ found in Na$_2$C$_{60}$. When looking at
Fig. \ref{T1K4NA2}, it is clear that a contribution of the order of 1~sec$%
^{-1}$ would not severely affect our discussion of K$_4$C$_{60}$.

The width of the peak depends on the variation of $\tau $ with
temperature and an Arrhenius law $\tau =\tau _0\exp (E_a/T)$ is
usually used, where E$_a$ is the activation energy for the
molecular motion and 1/$\tau _0$ the attempt frequency. These
parameters are probably similar for different compounds and a
typical width of about 50 K can be deduced from Fig.~\ref
{T1motion}. Interestingly, we have found a significant difference
between two Na$_2 $C$_{60}$ samples, called (1) and (2) on
Fig.~\ref{T1motion}. The results presented here are from sample
(1), sample (2) exhibits a similar behavior but seems to be of
somewhat poorer quality, as can be seen by the slightly shorter
T$_1$ values (see inset of Fig. \ref{T1motion}) or slightly
broader linewidth. The molecular motion peak, although clearly
present, is broader in sample (2), which is likely due to a
distribution of the motion parameters, and consequently its
maximum amplitude is smaller. Data on Fig. \ref{T1K4NA2} imply
that, if present, the peak in K$_4$C$_{60}$ must be very broad,
which could be due to sample quality or intrinsic disorder of the
bct phase. Its intensity would be accordingly reduced, so that its
contribution is furthermore negligible.

\subsubsection{High temperature behavior}

The $1/T_1=cst$ law observed at high temperatures in K$_4$C$_{60}$
is then intrinsic. It is somewhat unusual as most relaxation
mechanisms give an increasing relaxation rate with increasing
temperature. Such a flat behavior is reminiscent of the relaxation
observed in dense paramagnets, which is caused by a coupling to
localized paramagnetic centers. Within our model of
singlet-triplet excitations of JTD balls, we do have such centers
at high temperatures, namely the triplet states. Nevertheless, our
first expectation would be to observe such a law only when the
population of these levels saturates, for temperatures above the
ST gap, i.e. T~$>800~K$. More correctly, as 1/T$_1$ measures the
imaginary part of the electronic susceptibility{\it \ }(see Eq.
3), it is sensitive to both the {\it nature} and the {\it dynamic}
of the relevant electronic excitations. In our case, this means
that both the number and the lifetime of the triplet states
contribute to 1/T$_1,$ so that an abrupt change in the temperature
dependence of one of this quantity could explain the change in 1/T$_1$. In Na%
$_2$C$_{60}$, we observe a similar deviation from the activated law before
the expected saturation for $T=Eg,$ although it is not constant like in K$_4$%
C$_{60},$ but keeps increasing slightly up to 700 K.\ As the deviation is
present in both systems, it must contain some insights of their physics.

The first possibility is that the activated law fails to describe
the data over the full temperature range because there are other
thermally accessible excitations, like for example the singlet
state of JTD 2 in Fig. \ref {molecule}, if J is small enough. A
variant of this idea is that the arrangement of the molecular
levels could be modified with increasing temperature. JTD being
sensitively coupled to the structure's crystal field, it is likely
that even small structural modifications could affect the
equilibrium between different JTD. A second possibility that goes
beyond this ``molecular approach'', is related to the fact that we
deal here with solids that are very close to a metal-insulator
transition, where hopping is certainly not strictly forbidden. The
introduction of an hopping term in the JT hamiltonian mixes
different molecular states and can also affect the nature of the
ground state. An estimation of the temperature dependence of the
lifetime of the triplet state is obviously a complicated problem,
as it probably involves both the dynamic of the Jahn-Teller
distortion and hopping rates as a function of temperature.
Nevertheless, if a static distortion is for example stabilized at
low temperatures, allowing eventually the development of a
cooperative distortion, it could certainly affect 1/T$_1$.

In any case, the evolution of 1/T$_1$ at high temperature cannot be
understood with a model of isolated molecules and forces us to take into
account interactions between the balls and/or with the structure. For Na$_2$C%
$_{60}$, it is for example suggestive that 1/T$_1$ departs from
the activated behavior near the temperature of the structural
orientational transition taking place at 310 K \cite{Yldirim}. In
an attempt to identify the origin of the change in 1/T$_1$, we now
turn our attention to details of the structure that can be studied
by NMR to see whether there is any detectable change in the
corresponding temperature range.

\section{Interplay between JTD and local symmetry}

There are very few cases where Jahn-Teller distortions have been
observed directly in fullerides. One example is
C$_{60}$-tetraphenylphosphonium bromide, a salt where C$_{60}^{-}$
molecules are well separated from each other. A cooperative
Jahn-Teller state is thought to develop below 120 K, because a
splitting of the Lande factor has been observed by ESR below 120 K
\cite{Penicaud}. Interestingly, a transition to an orientationally
ordered state is observed at about the same temperature
\cite{Launois}, so that it seems likely that the new orientational
order stabilizes the collective distortion. This example motivates
us to relate to structural modifications the particular evolution
of 1/T$_1$ at high temperatures. NMR, and especially alkali NMR,
has proved to be a very sensitive probe for small structural
distortions in fullerides \cite{Pennington}. We first focus on the
effect of the structural transition in Na$_2$C$_{60}$ on the
electronic properties, as seen by $^{23}$Na NMR. We then review
other signs of structural evolution in Na$_2$C$_{60}$ and
K$_4$C$_{60}$ that indeed seem to coincide with the change in the
electronic behavior.

\subsection{Structural transition in Na$_2$C$_{60}$ studied by $^{23}$Na NMR}

The analysis of the electronic properties of alkali fullerides
might be complicated by the structural modifications associated
with orientational ordering or freezing of the molecular motions.
Although such structural modification could appear minor at first
sight, it has for example often be argued in the case of n=3 that
$sc$ phases behave quite differently from $fcc $ phases
\cite{Yldirimscfcc}. In the case of Na$_2$C$_{60}$ the occurrence
of an orientational transition at 310 K might then affect deeply
the electronic properties. At least, it seems reasonable to assume
that some parameters (such as the gap value) could be different on
both sides of the transition. Fitting the temperature dependence
of $\chi $ or 1/T$_1$ then becomes more difficult.

\begin{figure}[tb]
\centerline{ \epsfxsize=0.45\textwidth{\epsfbox{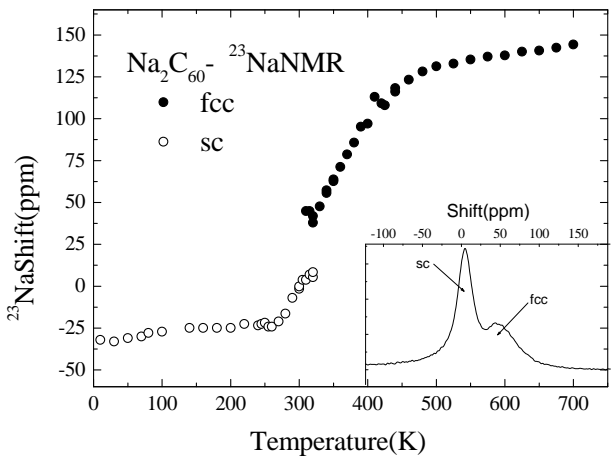}} }
\caption{$^{23}$Na NMR shifts (with respect to NaCl) in
Na$_{2}$C$_{60}$ for the $fcc$ phase (solid circles,
T\mbox{$>$}310K) and the $sc$ phase (open circles). Inset :
$^{23}$Na spectrum at 310 K showing the coexistence of the two
structures.} \label{Na2shift}
\end{figure}

This can be clarified by $^{23}$Na NMR because $^{23}$Na spectra
differ in the two phases, which allows to discriminate what
happens in the $sc$ and $fcc$ phases respectively. In the inset of
Fig. \ref{Na2shift}, a spectrum at 310 K shows the coexistence of
the two phases, which we observe from 300 K to 315 K. Following
the shift of each line, as done in Fig. \ref{Na2shift}, we can
extract the susceptibility in both phases independently and
observe that $\chi $ increases in {\it both} $sc$ and $fcc$
phases. We also clearly see that a ``saturation'' appears {\it in
the $fcc$ phase} at 400K, and not at the
structural transition. This contradicts the impression given by $^{13}$C 1/T$%
_1,$ which seems to change at the structural transition, but this
is more precise as $^{13}$C NMR does not resolve two different
signals and only the average value is accessible.

\begin{figure}[tb]
\centerline{ \epsfxsize=0.45\textwidth{\epsfbox{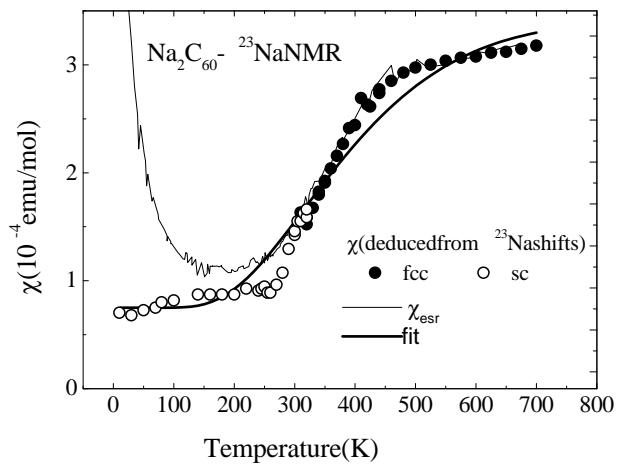}} }
\caption{ ESR susceptibility measured on the same Na$_{2}$C$_{60}$
batch than the NMR sample (thin line). Symbols represents an
extrapolation of the true susceptibility extracted from $^{23}$Na
NMR shifts as explained in the text. The thick line is a fit to
the model described in the text (Eq. 5). } \label{Na2ESR}
\end{figure}

In the $fcc$ phase, the scaling between $K$ and the ESR
susceptibility is excellent (see Fig. \ref{Na2ESR}) and the
hyperfine coupling $A_{fcc}$ can be extracted using Eq. 1,
provided that $\sigma _{fcc}$ is temperature independent (which is
a
usual assumption). We obtain $\sigma _{fcc}=-65$ ppm and $A_{fcc}=3500$ $%
Oe/\mu _B$. The evolution of the susceptibility in the $sc$ phase
is especially interesting as the ESR susceptibility is masked by a
large Curie term below 200 K. Following Eq. 1, the discontinuity
at the transition could be attributed either to a different
hyperfine coupling $A$ or to a different susceptibility. The
hyperfine coupling is defined by the local environment of a Na
atom, which is indeed very different in $sc$ or $fcc$ phases. In
the {\it fcc} phase, the orientations of the four neighboring
C$_{60}$ balls are such that Na faces four hexagonal rings,
whereas, in the $sc$ phase, it faces only one hexagonal ring and
three double bonds (see ref. \cite {PrassidesScience} and picture
of Fig. \ref{structuresc}). Therefore, there are probably two
different hyperfine couplings $A_{sc}$ and $A_{fcc}$. As ESR or
$^{13}$C 1/T$_1$ do not exhibit any obvious discontinuity at the
transition, $\chi $ is more likely to be continuous. Assuming that
$\sigma $ and $\chi $ do not change at the transition, we find
$A_{sc}$ = 2300 Oe/$\mu _B$ and the variation of $\chi $ deduced
from NMR is reported on Fig.~\ref {Na2ESR} by open circles for
$sc$ phase.

We can now compare this refined estimation of $\chi $ to the
singlet-triplet model. A first conclusion is that $\chi $ tends to
a constant value at low temperatures corresponding to $\chi
=7.10^{-5}$ emu/mol. Although, this is somehow dependent on our
assumption that $\sigma _{sc}=\sigma _{fcc},$ we have found a very
similar value (6 10$^{-5}$ emu/mol$)$ when trying to compare
directly T$_1$ and $\chi _{esr}$ in ref. \cite{BrouetPRL2001},
which gives some confidence in this estimate. This suggests the
presence of a Pauli-like contribution in Na$_2$C$_{60}$, which
likelihood will be discussed in the last section of this paper.

For the activated part, the singlet-triplet model of Fig.~\ref{molecule}
predicts a susceptibility $\chi (T)$~$=~n(T)~*~\chi (S_1),$ where $n(T)$ is
the number of thermally populated triplet states and $\chi (S_1)$ the Curie
susceptibility for triplet states.

\begin{equation}\label{chi}
\chi (T)=\frac{3\exp (-\Delta /T)}{2+3\exp (-\Delta /T)}*\frac{8\mu _B^2}{%
3k_BT}
\end{equation}

The thick line on Fig. \ref{Na2shift} is a fit to such a relation with $%
\Delta =100$ meV, which is 25 \% smaller than by using the
estimation based on 1/T$_1$ data below room temperature.
Quantitatively, the measured susceptibility corresponds to 80~\%
of that estimated by Eq. 5, which sounds reasonable. Although this
law clearly captures much of the physics of this phase, it does
not fit satisfyingly our data over the whole $T$ range. We observe
a steeper increase of the susceptibility at 250 K and a larger
inflexion at 450 K. This suggests that ``something else'' might
come into play at these temperatures, that helps or hinders the
population of triplet states, and that it is not purely a thermal
process. One of the possibility that comes into mind is a small
structural modification that would stabilize a particular state.

\subsection{Quadrupole effects on $^{23}$Na in Na$_{2}$C$_{60}$}

If the orientational structural transition does not seem to modify deeply
the behavior of Na$_2$C$_{60},$ we present here the observation of further
changes in local symmetry happening in the $sc$ phase that could couple to
Jahn-Teller distortions and interact with the electronic properties.

\begin{figure}[tb]
\centerline{ \epsfxsize=0.45\textwidth{\epsfbox{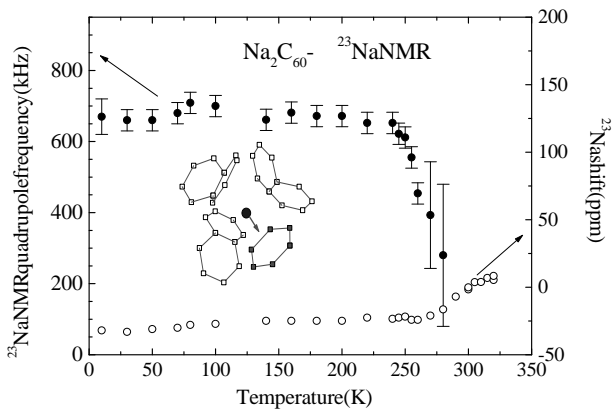}} }
\caption{Structural distortion in Na$_{2}$C$_{60}$ evidenced by
the appearance of a nuclear quadrupole frequency in the $^{23}$Na
spectra below 250 K (solid points, right scale). The picture
suggests that the distortion is due to a displacement of Na
towards the hexagon of one of its four C$_{60}$ neighbor. Left
scale : $^{23}$Na NMR shift that starts increasing when the
quadrupole frequency disappears. } \label{structuresc}
\end{figure}

As $^{23}$Na is a spin 3/2, it is sensitive to electric field
gradients (EFG) arising from deviations from cubic symmetry at the
Na site. In the {\it fcc} phase, there are no detectable
quadrupole effects, as expected in the cubic environment of one
tetrahedral site. However, in the{\it \ sc} phase, a decrease in
the NMR intensity at the $fcc$ to $sc$ transition indicates that
quadrupole effects are present. In this case, the
(-3/2$\rightarrow -1/2)$ and (1/2$\rightarrow 3/2)$ nuclear
transitions are wiped out of the spectrum and only the central
nuclear transition (1/2~$\rightarrow $~-1/2) is detected. Below
200 K, a splitting of this central transition, which is
characteristic of second order broadening by the EFG \cite
{Abragam} is detected. The (1/2$\rightarrow -1/2)$ line can be
fitted at 100 K by an EFG\ tensor with a quadrupole frequency $\nu
_q=700$ $kHz$ and a small asymmetry $\eta =0.2$
\cite{Kirchberg00}$.$ The evolution of $\nu _q$ with temperature
is displayed on Fig. \ref{structuresc}, it shows that this
quadrupole effect progressively increases when the temperature is
lowered from 280 to 230$~$K.

Where does this EFG come from ? As represented on Fig.
\ref{structuresc}, in the $sc$ phase, the environment of $^{23}$Na
is quite asymmetric. This could favor a displacement of Na along
the cube diagonal (indicated by the arrow) towards the hexagonal
ring, as was observed by x-ray in the structurally similar
Na$_2$CsC$_{60}$ \cite{PrassidesScience}, that would create an
electric field gradient. C$_{60}$ molecular motions have to be
reduced to allow this displacement. More precisely, rotation
around one axis could still exist (and it probably persists down
to about 180 K where we observe the peak in $^{13}$C NMR 1/T$_1$)
but reorientation of the rotation axis must be nearly prohibited
(such a decomposition of the molecular motion was proposed for
K$_3$C$_{60}$ \cite{Yoshinari}$).$ Here, we believe that this
``slow'' reorientational motion is slowing down progressively when
we begin to observe static quadrupole effects around 280 K and is
totally frozen below 250 K.

It is quite striking that the $^{23}$Na shift, plotted again on
Fig. \ref {structuresc} for comparison, starts to increase just as
the EFG disappears and this strongly suggests a relation between
the two effects. One possibility is that this increase reflects
the one of the hyperfine coupling as the Na atom moves because of
the distortion. However, we have seen in the previous section that
the shift can be rather well understood in terms of a
singlet-triplet susceptibility, so that we believe that it is the
susceptibility that starts to increase suddenly when the
distortion disappears. This indeed would explain the steeper
increase of the shift compared to the singlet-triplet model noted
on Fig \ref{Na2ESR}. This suggests that the distortion stabilizes
the singlet state and that triplet states can only be
significantly populated when it disappears. The exact microscopic
origin of such an interplay is not yet clear. The orbital moment
of the triplet distortion might be incompatible with the crystal
field induced by the structural distortion, which forbids their
existence.

\subsection{Change of the symmetry of the C$_{60}$ molecular motion in K$_4$C%
$_{60}$}

\begin{figure}[b]
\centerline{ \epsfxsize=0.45\textwidth{\epsfbox{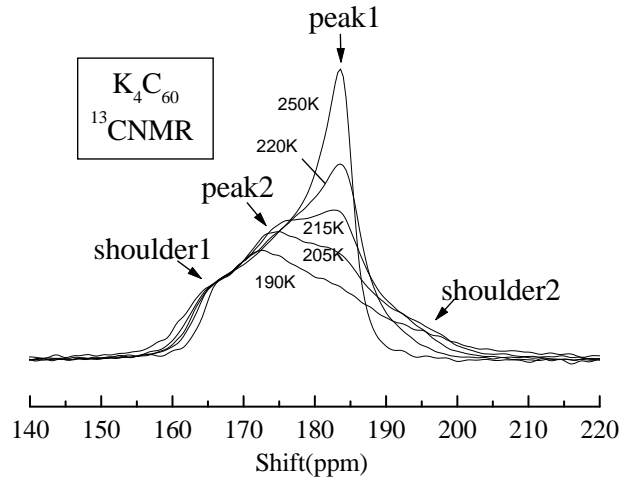}} }
\caption{$^{13}$C NMR spectra in K$_{4}$C$_{60}$ from 190 K to 250
K showing a change in the $^{13}$C lineshape.} \label{K4shoulder}
\end{figure}

\begin{figure}[tb]
\centerline{ \epsfxsize=0.45\textwidth{\epsfbox{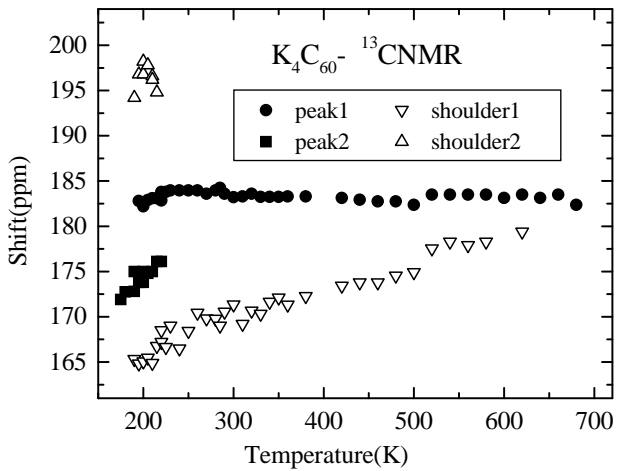}} }
\caption{Shifts of the different peaks indicated on Fig. 10 in the
$^{13}$C NMR spectrum of K$_{4}$C$_{60}$ from 200 to 700~K. }
\label{K4shoulderb}
\end{figure}

In K$_4$C$_{60},$ there is no ordering transition but we report
here in Fig. \ref{K4shoulder} modifications of the $^{13}$C
lineshape that indicate a structural evolution. At high
temperatures, the $^{13}$C spectrum consists of one narrow
symmetric line, as expected because of the motional narrowing of
the spectrum. However, a shoulder appears below 580 K on the low
frequency side, and becomes progressively better defined as the
temperature is lowered. This lineshape is characteristic of a
small axial anisotropy and probably corresponds to the
developement of the uniaxial motion, which would not be completely
averaged anymore by molecular reorientations. Note that this
anisotropy is however still very small with respect to the low
temperature one (200 ppm). Below 250 K, the spectral weight is
quite suddenly transferred to the right of the spectra, as noted
on the figure by the appearance of a ``peak 2'' and ``shoulder
2''. Fig. \ref {K4shoulderb} summarizes the situation by
displaying the shift of the various peaks as a function of
temperature. This complex behavior is not understood, but it is
probably related to a change in the symmetry of the molecular
motion. This shows that, even though there is no reported
structural transition in K$_4$C$_{60}$ in this temperature range,
the local symmetry changes. Let us recall that a true structural
transition has been detected by differential thermal analysis in
K$_3$C$_{60}$ at 200 K that has never been associated so far to a
precise structural deviation \cite {Yoshinari}.

As a matter of fact, 1/T$_1$ saturates precisely in this
temperature range, as can be checked on Fig. \ref{T1K4NA2}. This
``coincidence'' bears some similarity with the case of
Na$_2$C$_{60},$ which incites us to take it seriously. When
discussing the 1/T$_1$=cst regime, we mentioned that it could be
understood by a coupling to a fixed number of triplet states. This
would require that the structural change at 250 K favors the
triplet states almost exclusively.

Very recently an infrared study of K$_4$C$_{60}$ has revealed a
splitting of the two high frequency T$_{1u}$ modes from a doublet
above 250 K to a triplet at lower temperature \cite{Kamaras}. This
observation definitely establishes a symmetry breaking in this
temperature range, although the exact nature of the JTD and their
ordering in the high and low temperature phases is still unclear.

\section{Coexistence with a metallic character ?}

So far, our study has been mainly focused on the molecular singlet-triplet
excitations that dominate the high temperature behavior. However, we have
seen that a Pauli-like contribution seems to be present in $\chi $ for Na$_2$%
C$_{60},$ which could imply that Na$_2$C$_{60}$ is weakly metallic. Such a
possible coexistence of band-like excitations with typically molecular ones
is an important issue. It could indeed help to clarify the nature of the
metal-insulator transition in A$_n$C$_{60}$. Two different kind of such
transitions could be considered, the one that takes place in Rb$_4$C$_{60}$
as a function of lattice spacing, which has been observed under applied
pressure \cite{Kerkoud} and the one expected as a function of doping if one
could go continuously from Na$_2$C$_{60}$ to A$_3$C$_{{60}}$ to A$_4$C$_{60}$%
. Therefore, we pay hereafter particular attention to this subject by
studying the low temperature behavior of 1/T$_1$ in Na$_2$C$_{60}$, which
probes the nature of the excitations of the ground state of this system.

\begin{figure}[b]
\centerline{ \epsfxsize=0.45\textwidth{\epsfbox{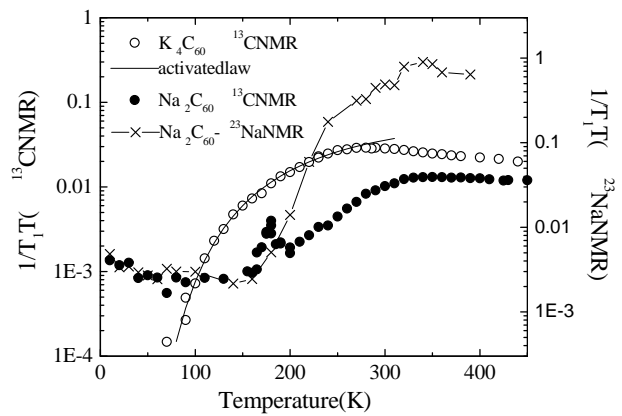}} }
\caption{ Comparison of 1/T$_{1}$T for $^{13}$C and $^{23}$Na in
Na$_{2}$C$_{60}$ and $^{13}$C in K$_{4}$C$_{60}$ as a function of
temperature. Note the logarithmic scale. The solid line is the
activated law of Fig.~5 for K$_{4}$C$_{60}$.} \label{NaCar}
\end{figure}

The temperature dependence of 1/T$_1$T at low T in Na$_2$C$_{60}$ and K$_4$C$%
_{60}$ is emphasized in the logarithmic plot of Fig.~\ref{NaCar}.
It can be seen that in Na$_2$C$_{60}$, 1/T$_1$ deviates from the
activated behavior
below 100 K, actually 1/T$_1$T tends to a constant value for $^{13}$C and $%
^{23}$Na. On the other hand, in K$_4$C$_{60},$ 1/T$_1$T follows
the activated law of Fig. 5 down to the lowest measured
temperature. The very long value for T$_1$ in this system at this
temperature prevents us to study this behavior further. In
Rb$_4$C$_{60}$ also, the low temperature data do not follow the
activated behavior but was ascribed to a much smaller gap (10 meV)
\cite{Kerkoud}. As 1/T$_1$T = cst is the Korringa law expected in
a metal, this reinforces the possibility of a weak metallicity of
Na$_2$C$_{60}$. Before concluding firmly on this eventuality, we
have to consider possible complications in the interpretation of
the relaxation data.

\subsection{Relaxation curves}

\begin{figure}[tb]
\centerline{ \epsfxsize=0.45\textwidth{\epsfbox{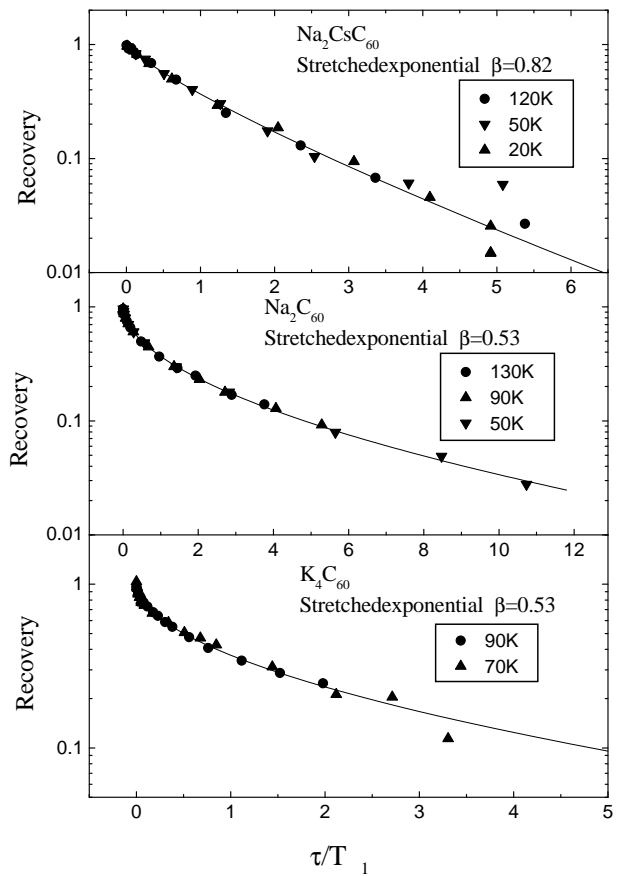}} }
\caption{ Recovery curves for $^{13}$C in Na$_{2}$CsC$_{60}$,
Na$_{2}$C$_{60} $ and K$_{4}$C$_{60},$ for different temperatures
(see legend). [ M($\protect\tau $) -M$_{0}] $ $/$ M$_{0}$ is
plotted on a logarithmic scale$,$ where M($\protect\tau )$ is the
echo intensity at a delay $\protect\tau $ after the saturation
pulse and M$_{0}$ the intensity at saturation ($\protect\tau
\rightarrow \infty $). The recovery curves are scaled for the
different temperatures by normalizing $\protect\tau $ with the
value of T$_{1}$ extracted from a stretched exponential fit with
an exponent $\protect\beta $ indicated on the graph. The solid
line is a fit to such a recovery law. } \label{recovery}
\end{figure}

To be sure of the temperature dependence of 1/T$_1$ down to the lowest
temperature, one has to check first that the shape of the nuclear
magnetization recovery curve after saturation is not changing. This question
is not trivial in fullerides as a non-exponentiality always appears at low
temperatures, giving some ambiguity in the actual definition of an average T$%
_1$ value. In Na$_2$C$_{60}$ and K$_4$C$_{60}$, this happens below 150 K,
but Fig. \ref{recovery} shows that the relaxation curves can be {\it %
scaled together} for different temperatures. This means that they keep a
similar shape with varying temperature within experimental accuracy, and
therefore a single time parameter which is taken as T$_1$ can be used to
characterize the variation of the relaxation below 150 K. The actual value
of T$_1$ somehow depends on the expression used to fit the magnetization
recovery curve but not its temperature dependence.

In Na$_2$C$_{60}$, the relaxation curves can be described by a
stretched exponential M(t) = exp((-t/T$_1$)$^\beta $) with values
of $\beta $ ranging from 0.5 to 0.6. Within experimental accuracy,
many different laws could describe this dependence, from
multi-exponential recovery (2 sites or more) to stretched
exponential. We choose the latter not for physical reasons but
because it allows to compare easily the relaxation in different
systems. The same fit applies in K$_4$C$_{60},$ although the
experimental accuracy is not sufficient to determine $\beta $ very
precisely. All the curves presented on Fig. \ref{recovery} are
fitted with $\beta =0.53$ to illustrate the adequation of this
fitting function. For comparison, the recovery curves in
Na$_2$CsC$_{60}$ are also plotted, they are typical of recovery
curves found in A$_3$C$_{60}$ compounds \cite{Holczer,Maniwa}.
Here, the deviation from exponentiality is smaller, as illustrated
by the value of the exponent for the stretched exponential ($\beta
=0.82$) closer to unity. The difference found for the exponent
$\beta $ for the two classes of systems indicates a significant
difference in the relaxation curves and one can wonder whether it
has a physical meaning.

In A$_3$C$_{60},$ the general belief \cite{Holczer,Maniwa} is that the
widely observed non-exponentiality is due to the differentiation between
three slightly inequivalent $^{13}$C sites on one C$_{60}$ ball, when the
motion of the balls is frozen. As the structure is much more different
between Na$_2$C$_{60}$ and K$_4$C$_{60}$ than Na$_2$C$_{60}$ and Na$_2$CsC$%
_{60},$ it seems unlikely that the change of relaxation behavior
has a structural origin. We would rather suggest that it is
related to a different nature of the relaxation in K$_4$C$_{60}$
and Na$_2$C$_{60},$ more localized on the ball. Alternatively,
this could be due to the addition of an extrinsic term caused by
paramagnetic impurities.

\subsection{Role of impurities in the low T relaxation}

In insulating solids, as the intrinsic T$_1$ becomes long at low
temperature, even a small number of paramagnetic impurities could
become a dominant relaxation process. As a matter of fact, a
rather high concentration of paramagnetic impurities seems to be
always present in these compounds and their role in the low T
relaxation must be considered. Paramagnetic impurities would
likely produce a saturation of 1/T$_1$ at low temperatures, i.e.
an {\it increase} of 1/T$_1$T \cite{Abragam}. One could imagine
that this ``compensates'' the decrease of the singlet-triplet
component to mimic a 1/T$_1$T=cst law.

The comparison between Na$_2$C$_{60}$ and K$_4$C$_{60}$ can help
to test whether this is likely. The paramagnetic contribution
should have the same characteristics in both compounds but be
proportional to the number of paramagnetic impurities in a given
sample. We have characterized the paramagnetic impurity content by
ESR on samples issued from the same batches as the NMR ones. From
the low temperature Curie tail observed by ESR, the impurity
concentrations are estimated to be about 2\% per C$_{60}$ in
Na$_2$C$_{60}$ and 1\% in K$_4$C$_{60}$. We do observe longer
T$_1$ in K$_4$C$_{60}$, but they are too long (by already a factor
10 at 70 K, as can be seen on Fig. \ref{NaCar}) to be explained by
a difference of a factor 2 in impurity content. Data in
K$_4$C$_{60}$ allow to set a maximal value for the contribution of
impurities in Na$_2$C$_{60}$ to 1/T$_1$T$<3$ 10$^{-4}$~($\sec$
K)$^{-1}$ at 70 K, which is too small to be responsible for the
deviation from the activated law.

Another independent indication of the intrinsic character of the
relaxation comes from the fact that it is also observed by
$^{23}$Na NMR. As the increase of 1/T$_1$ between 150 K and 300 K
has very different magnitudes for $^{13}$C and $^{23}$Na, probably
because a quadrupole term is present in $^{23}$ Na 1/T$_1,$ it
seems impossible that an intrinsic and extrinsic term could
compensate at low T to give an identical temperature dependence on
the two nuclei over a temperature range as large as 150 K.

{\it We then conclude that our experiment probes a metallic character in Na$%
_2$C$_{60}$.} As a proof of consistency, the value of n(E$_f$)
needed to explain the 1/T$_1$T=cst law by the Korringa mechanism
can be estimated to be 1eV$^{-1},$ which agrees with the value of
$\chi \approx 7.10^{-5}$~emu/mol found by $^{23}$Na NMR
\cite{BrouetPRL2001}.

\section{Conclusion}

In conclusion, the main properties of Na$_2$C$_{60}$ and
K$_4$C$_{60}$ appear to be similar, which shows that they
represent the typical behavior of two electrons or two holes in
the t$_{1u}$ band. Molecular excitations dominate their
properties; starting from a singlet ground state, excitations to a
triplet one are thermally accessible. We attribute the existence
of these two states to the two most stable Jahn-Teller
distortions, which have respectively a singlet and triplet ground
state. We further show that this molecular approach is
insufficient to describe the full temperature range. $\chi $ and
1/T$_1$ seem to change more suddenly that one would expect in a
completely thermal process. The structural transition from $fcc$
to $sc$ is not found essential in the variation of the properties
of Na$_2$C$_{60}$. However, in both compounds, we evidence changes
in the structure concomitant with the change in 1/T$_1$, probably
associated with the slowing down of molecular motions. This
suggests that there is a coupling between the structure and the
stabilization of JTD, maybe due to crystal field effects acting on
the orbital moment of the different Jahn-Teller configurations.

We also give evidence that the singlet states are not strictly
localized on one molecule. This is most clear by the low energy
excitations observed in Na$_2$C$_{60}$ at low temperatures, which
are best described by a metallic-like process with a small value
of the density of states. Interactions between the balls also
renormalize the value of the ST gap and yield to deviations with
respect to a purely molecular model in the behavior of 1/T$_1$ and
$\chi $.

The interplay between molecular and band-like properties is
certainly one of the most intriguing feature of fullerides and a
good understanding of this effect is a prerequisite before
addressing the case of A$_3$C$_{60}$. The fact that these
superconducting fullerides are surrounded by almost insulating
phases has not always received much attention, probably because it
is not easily possible to go from one phase to the other. When
transport is measured as a function of alkali doping, usually on
thin films or with the recently synthesized FET devices
\cite{Batlogg}, no sharp metal to insulator transitions can be
observed, which gives the impression that a rigid band filling
picture could be applied. However, little is known on the
homogeneity of the stoichiometries of these films and phase
separation could take place there, like they do for bulk
compounds. The well established fact that superconductivity is
restricted to a very limited doping range around n=3 is a strong
deviation from the expectations of a BCS theory usually applied to
these materials. This indicates particular properties for the {\it
integer filling} n=3. This might also be true for n=2 and 4, and
the metallic state could be nearly suppressed {\it only} in these
two cases. This importance of integer fillings might be related to
the stronger correlation effects that appear in this case combined
with the possibility of stabilizing molecular JTD. On one hand,
the case of Na$_2$C$_{60}$ studied here indicates that metallic
and molecular properties are not exclusive but {\it do} coexist.
On the other hand, the study of CsC$_{60}$ that we will present in
paper II shows that JTD C$_{60}^{2-}$ have an enhanced stability
in this compound with nominally one electron per C$_{60}$. We will
develop in paper III the idea that the remarkable properties of
A$_3$C$_{60}$ are due to an optimum cooperation between metallic
and molecular aspects, because of its symmetric position between
n=2 and 4.

\end{document}